\documentclass[12pt,a4paper]{article}
\pdfoutput=1
\usepackage{jheppub}
\usepackage{amsfonts,amsmath,amssymb}
\usepackage{enumerate}
\usepackage{hyperref}
\usepackage{graphicx}

\newcommand{\be}{\begin{equation}}
\newcommand{\ee}{\end{equation}}

\newcommand{\bea}{\begin{eqnarray}}
\newcommand{\eea}{\end{eqnarray}}

\newcommand{\bes}{\begin{subequations}}
\newcommand{\ees}{\end{subequations}}

\newcommand{\nn}{\nonumber}

\newcommand{\ra}{\rightarrow}

\newcommand{\Tr}{\mbox{Tr}}

\newcommand{\cp}{\mathbb {CP}}

\newcommand{\kp}{k_{+}}
\newcommand{\km}{k_{-}}

\title{Operator counting for ${\cal N}=2$ Chern-Simons gauge theories with 
chiral-like matter fields}
\author{Hyojoong Kim} 
\author{and Nakwoo Kim}
\affiliation{Department of Physics
and Research Institute of Basic Science, \\ Kyung Hee University, 
 Hoegi-dong, Dongdaemun-gu,
 Seoul 130-701, Korea}

\emailAdd{h.kim@khu.ac.kr}
\emailAdd{nkim@khu.ac.kr}

\abstract{The localization formula of Chern-Simons quiver gauge theory on $S^3$ nicely reproduces the geometric data such as volume of Sasaki-Einstein manifolds in the large-$N$ limit, at least for vector-like models. The validity of chiral-like models is not established yet, due to technical problems in both analytic and numerical approaches. Recently Gulotta, Herzog and Pufu suggested that the counting of chiral operators can be used to find the eigenvalue distribution of quiver matrix models. 
In this paper we apply this method to some vector-like or chiral-like quiver theories, including the triangular quivers with generic Chern-Simons levels which are dual to in-homogeneous Sasaki-Einstein manifolds $Y^{p,k}(\mathbb{CP}^2)$. The result is consistent with AdS/CFT and the volume formula. We discuss the implication of our analysis.  }
\begin{document}
\maketitle
%\arxivnumber{}
%\begin{document}
%\maketitle

%%%%%%%%%%%%%%%%%%%%%%%%%%%%%%%%%%%%%%%%%%
%%%%%%%%%%%%%%%%%%%%%%%%%%%%%%%%%%%%%%%%%%
\section{Introduction}

There has been a remarkable progress in our understanding of the low energy dynamics of multiple M2-branes in recent years. A three-dimensional ${\cal{N}}=6$ Chern-Simons-matter theory with $U(N)_k \times U(N)_{-k}$ gauge group with four bifundamental chiral multiplets was proposed by Aharony, Bergman, Jafferis and Maldacena (ABJM) \cite{Aharony:2008ug} as the theory of M2-branes probing $\mathbb{C}^4/\mathbb{Z}_k$. Its relation to M-theory in $AdS_4\times S^7$ in light of
 AdS/CFT correspondence has been the theme of many papers since then, see e.g. \cite{Klebanov:2009sg} and references therein.
One of the most impressive tests is the computation of partition function for the ABJM model on $S^3$ \cite{Kapustin:2009kz}\cite{Drukker:2010nc} 
using the localization technique \cite{Pestun:2007rz}. It has been checked that the free energy exhibits the scaling behavior $N^{3/2}$ in agreement with the prediction from 
M-theory \cite{Drukker:2010nc}\cite{Klebanov:1996un}. 
  
It is then natural to ask whether this program can be generalized to other AdS/CFT models with less supersymmetries. It turns out that ${\cal N}=2$ is the minimal amount of supersymmetry needed for localization technique \cite{Jafferis:2010un}\cite{Hama:2010av}. The dual supergravity geometry is $AdS_4\times Y_7$, where $Y_7$ should be a seven-dimensional Sasaki-Einstein  manifold. 
 There are now a number of different ${\cal{N}}=2$ Chern-Simons-matter theories which are proposed to be dual to M-theory on the $AdS_4 \times Y_7$ background with  a Sasaki-Einstein manifold $Y_7$
\cite{Martelli:2008si}\cite{Ueda:2008hx}\cite{Hanany:2008fj}\cite{Franco:2008um}\cite{Franco:2009sp}\cite{Aganagic:2009zk}.
%Following the localization method in 4 dimensions \cite{Pestun:2007rz}, the partition function  of ${\cal{N}}\ge 3$ Chern-Simons gauge theories on $S^3$ can be reduced to the multi-matrix model \cite{Kapustin:2009kz}. And finally the $N^{3/2}$ scaling behavior of the M2-branes degrees of freedom was explained for the ABJM model \cite{Drukker:2010nc}. For the ${\cal{N}}=3$ theories the partition function was calculated by using the saddle point approximation at large $N$ limit \cite{Herzog:2010hf}. Extremizing the free energy they obtained $N^{3/2}$ scaling and reproduced the volume of the tri-Sasakian manifold.
When a three-dimensional theory has ${\cal{N}}=2$ supersymmetries, the conformal dimensions (or equivalently the R-charge) of chiral fields may differ from the canonical one. 
%and the localization prescription was appropriately generalized in \cite{Jafferis:2010un}\cite{Hama:2010av}.
It is also proposed that the exact R-charges at IR fixed point can be determined by extremizing the free energy $F= - \ln Z$ \cite{Jafferis:2010un}. This conjecture then leads 
to the F-theorem \cite{Jafferis:2011zi}\cite{Amariti:2011da}\cite{Amariti:2011xp}\cite{Klebanov:2011gs} that the free energy on $S^3$ decreases along the RG flow and must be stationary at fixed point.  Using the saddle point approximation the partition function was calculated in the large-$N$ limit for various
 models \cite{Martelli:2011qj}\cite{Cheon:2011vi}\cite{Jafferis:2011zi}, following an 
earlier work on ${\cal N}=3$ models in \cite{Herzog:2010hf}. For the models studied, which are all {\it vector-like}, 
the computed free energy again exhibits the required $N^{3/2}$ scaling and reproduces  exactly the volume of the dual Sasaki-Einstein 7-manifold from the coefficient.

It turns out that for {\it chiral-like} models the correspondence is much more delicate. For such models the quiver diagram is not invariant under
conjugation, and in particular the partition function is not real-valued.
 For them the long range forces between the eigenvalues do not cancel and the free energy is apparently proportional to $N^2$ \cite{Jafferis:2011zi}, 
 instead of $N^{3/2}$. To overcome this technical difficulty,
 Amariti and Siani recently proposed a symmetrization technique in  \cite{Amariti:2011jp}. In the computation, one considers a symmetrized form of the  integrand which effectively replaces a bifundamental field with a pair of {\it half} bifundamentals in mutually conjugate representations. Indeed, the volume of the $Q^{2,2,2}/\mathbb{Z}_k, M^{1,1,1}/\mathbb{Z}_k$ dual to chiral-like models was successfully reproduced in \cite{Amariti:2011uw}\cite{Gang:2011jj}. 
 %{\bf Please read this part carefully and tell me if this is correct or not.} 
 
 However this prescription is still limited in applicability. For the models investigated so far, the eigenvalue distribution is symmetric under $u\ra -u$ even though this is not an obvious symmetry of the integral. And in those models the R-charge of monopole operators do not make nontrivial contribution in the integrand, so we can set it to zero. As we will argue later, incorporating monopole R-charge is crucial when we extend to the inhomogeneous Sasaki-Einstein manifolds such as $Y^{p,k}(\mathbb{CP}^2)$. In fact one can easily see that the result from symmetrization does not work for the case of $Y^{p,k}(\mathbb{CP}^2)$. \footnote{ This is done by following the symmetrization prescription in \cite{Amariti:2011jp}\cite{Gang:2011jj} to write down the saddle point equations, and then making use of the general rule summarized in \cite{Jafferis:2011zi}.}
 At present it is not clear to us how to repair the matrix model for chiral-like models.

On the other hand, Gulotta, Herzog and Pufu discovered a relationship between the operator countings in the chiral ring of the gauge theory and the eigenvalue distributions of the matrix model with ${\cal{N}}=3$ supersymmetries \cite{Gulotta:2011si}. It is illustrated
 that this relation holds for ${\cal{N}}=2$ non-chiral gauge theories and
 the authors also provided prediction on the eigenvalue distribution of the chiral theory by counting the number of gauge invariant operators \cite{Gulotta:2011aa}. 
It was also generalized to ${\cal{N}}=3$ Chern-Simons theories with an ADE classification \cite{Gulotta:2011vp}\cite{Gulotta:2012yd}.  Readers are also referred to \cite{Berenstein:2011dr} for a more comprehensive study on the relation between operator counting and the dual geometry. 

In this note we posit that the information of the eigenvalue distribution of a given matrix model is encoded in the operator counting, and apply it to various ${\cal{N}}=2$ Chern-Simons theories which are {\it not} vector-like circular quivers. As a warm-up we study a non-chiral example of the dual ABJM model \cite{Hanany:2008fj,Franco:2008um}, which is vector-like but not a circular quiver. 
After constructing gauge invariant operators in terms of the bifundamental fields and the monopole operator, we count the number of operators for given R-charge and monopole number. Then we can obtain the eigenvalue distribution density function $\rho(x)$ and the imaginary part of the eigenvalues $y_a(x)-y_b(x)$, from the relation between the operator counting and the matrix model conjectured in \cite{Gulotta:2011si, Gulotta:2011aa}. As a consistency check, we calculate the volume of the 7-manifold and the 5-cycles and show that these volumes exactly agree with the geometric data. We  also study the two different chiral models dual to $AdS_4 \times Q^{1,1,1}/\mathbb{Z}_k$. The final example is Chern-Simons theory dual to $AdS_4 \times Y^{p,k}(\cp^2)$. We will show that the equation which extremizes the free energy with respect to the R-charges of the monopole operator gives exactly two cubic equations, just as presented in \cite{Martelli:2008rt}, which govern the geometry of $Y^{p,k}(\cp^2)$. This result, unlike the symmetrization prescription \cite{Amariti:2011uw}\cite{Gang:2011jj}, adds more credence to the chiral-like Chern-Simons model proposal and the operator counting prescription. 
We will make a short comment about the gauge theory dual to $Y^{p,k}(\cp^1 \times \cp^1)$.

This paper is organized as follows. In section \ref{review}, we briefly review the operator counting technique and the localization formula of ${\cal{N}}=2$ Chern-Simons gauge theories.
Section \ref{example} is the main part and we apply this operator counting method to the dual ABJM model and the chiral models dual to $Q^{111}/\mathbb{Z}_k$ and $Y^{p,k}(\cp^2)$.
%The geometry of the 7 dimensional inhomogeneous Sasaki-Einstein manifold $Y^{p,k}$ is summarized in the appendix \ref{SE7}. The particle solutions in $Y^{p,k}(\cp^2)$ is briefly reviewed in the appendix \ref{geodesic}.
We briefly discuss the implication of our results in Section \ref{discussion}.
%%%%%%%%%%%%%%%%%%%%%%%%%%%%%%%%%%%
\section{Reviews on Operator countings and matrix model}\label{review}
%%%%%%%%%%%%%%%%%%%%%%%%%%%%%%%%%%%%%%%%%%%%%%%%%
The holographic free energy for $N$ M2-branes with gravitational dual
$AdS_4\times Y_7$  is given as \cite{Klebanov:1996un}
\be
F=N^{3/2} \sqrt{\frac{2 \pi^6}{ 27 \textrm{Vol}(Y_7)}}.
\ee
Our goal in this paper is to see how this relation fares in various
examples of $AdS_4/CFT_3$, especially for chiral-like models.

Exact calculation of the partition function is a nontrivial task in principle,
but thanks to supersymmetry one can utilize the localization technique \cite{Pestun:2007rz}. 
Then the path integral is greatly simplified and one simply
 has an ordinary integration 
over eigenvalues of auxiliary $D$-field of gauge multiplets \cite{Kapustin:2009kz}.  
In the large-$N$ limit one can employ the saddle point method, and the integral is
determined basically by the eigenvalue dynamics on the complex plane. 
Thus the free energy can be written as a functional of $\rho(x)$ and $y_a(x)$. Here $\rho$ is the density of the eigenvalue distribution and $x(y_a)$ is the real(imaginary) part of the eigenvalue in continuum limit. It is convenient to introduce the Lagrange multiplier $\mu$ to impose the condition $\int d x \rho=1$. Then the free energy is given 
by extremizing with respect to $\rho(x), y_a(x)$ and $\mu$.

The matrix model integrand exhibits a number of  flat directions which are remnants of 
gauge invariance and the symmetry of quiver diagram \cite{Jafferis:2011zi}. An important point here is that the partition function is invariant under the shift of the R-charge, e.g.  $R[X_{ab}] \rightarrow R[X_{ab}]+\delta^{(b)}-\delta^{(a)}$ for bifundamental fields. 
Then only the R-charge of  gauge invariant operators (loops in the quiver diagrams) is of physical significance. We will frequently make use of this invariance to simplify the calculations in this paper. 

%Thus we cannot determine the R-charges of the individual field $X_{ab}$. Instead
%in our convention we vary the R-charge $\Delta_m$ and extremize the free energy with respect to $\Delta_m$. %{\bf maybe this part is wrong. need to be made more precise}

Recently an alternative interpretation of eigenvalue dynamics was given from the operator countings in the chiral ring 
\cite{Gulotta:2011si,Gulotta:2011aa,Gulotta:2011vp}.
Let $\psi (r,m)$ be the number of the gauge invariant operators with R-charges and the monopole charges less than $r, m$ respectively. And let 
$\psi_{X_{ab}}$ be the number of the operators which do not have the bifundamental field $X_{ab}$. Then it was proposed that the following relations hold:
\bea
\frac{\partial^3 \psi}{\partial^2 r \partial m}\bigg |_{m=r x/\mu}&=& \frac{r}{\mu} \rho(x),\label{opcrho} \\
\frac{\partial^2\psi_{X_{ab}}}{\partial r\partial m}\bigg |_{m=r x/\mu} &=& \frac{r}{\mu} \rho(x) [y_b (x)- y_a(x)+R(X_{ab})].\label{opccy}
\eea
 By counting the number of gauge invariant operators, one can easily read off the matrix model information which can be used to calculate the volume of the 7-dimensional internal space and 5-cycles in the dual geometry.
\bea
\textrm{Vol}(Y_7)&=&\frac{\pi^4}{24} \int d \hat{x} \hat{\rho}(\hat{x})\, ,  \label{7-vol}\\
\textrm{Vol}(\Sigma_{X_{ab}})&=&\frac{\pi^3}{4} \int  d\hat{x} \hat{\rho}(\hat{x})(\hat{y}_b(\hat{x})-\hat{y}_a(\hat{x})+R[X_{ab}])\, , \label{cycle-vol}
\eea
where
\be
\hat{x} =\frac{x}{\mu},~~~~~\hat{\rho}(\hat{x})= \frac{\rho{(x)}}{\mu},~~~~~\hat{y}_a(\hat{x})=y_a(x).
\ee
 Then the free energy of the theory is simply given as 
\be
F=\frac{4}{3}\pi N^{3/2}\mu.
\ee

%%%%%%%%%%%%%%%%%%%%%%%%%%%%%%%%%%%%%%%%%%

%%%%%%%%%%%%%%%%%%%%%%%%%%%%%%%%%%%%%%%%%%
\section{Operator counting for non-circular quivers}\label{example}
In this section we count the number of the gauge invariant operators of various ${\cal{N}}=2$ Chern-Simons-matter theories. First we construct  gauge invariant operators using  matter chiral multiplets and monopole operators. Then we have a relation between the number of various operators from gauge invariance. The total R-charge is given simply as the sum of the R-charges for constituent chiral scalar fields. Exploiting the flat directions of the matrix model, we can set the R-charges of all the bifundamental fields, which make a basic closed loop in the quiver, to be the same. 
%and assign R-charge $\Delta_1$ etc. 
As we are dealing with toric cases, counting the number of the operators with the R-charge $r$ and the monopole charge $m$ is reduced to calculating the area of the polygon in the case of large $r$. One can consult Appendix C of \cite{Gulotta:2011aa} for more detail. The number of operators which do not have the chiral field $X$ is obtained by calculating the length of edge. % Here F-term condition plays a important role. 
Then we have the density function and the imaginary part of the eigenvalues in the matrix model and express the volume of the internal 7-manifold and the 5-cycles in terms of the R-charges of the monopole operator $\Delta_m$ and the bifundamental field $\Delta_1$ etc. Extremizing the volume with respect to these R-charges will give the correct value which agrees with the geometric computation. We choose the convention where the bifundamental field $X_{ab}$ is in the $(\bar{N}_a,N_b)$ representations and the diagonal monopole operator has charge $(k_1, k_2, \dots,k_n)$ in $U(N)^n$ gauge groups. 

%%%%%%%%%%%%%%%%%%%%%%%%%%%%%%%%%%%%%%%%%
\subsection{A non-chiral model : dual ABJM}
%%%%%%%%%%%%%%%%%%%%%%%%%%%%%%%%%%%%%%%%%

As a first and simple example, we consider the dual ABJM model \cite{Hanany:2008fj,Franco:2008um}. This theory has $U(N)\times U(N)$ gauge groups with Chern-Simons levels $(-k,k)$ and 2 bifundamental fields and 2 adjoint fields $\phi_i$ at one of the $U(N)$ gauge groups. The superpotential of this theory is
\be
W= \textrm{Tr} (X_{12}X_{21}[\phi_1,\phi_2]) \, .
\ee
\begin{figure}
\centering
\includegraphics[scale=0.4 ,trim= 20 260 300 150,clip=true]{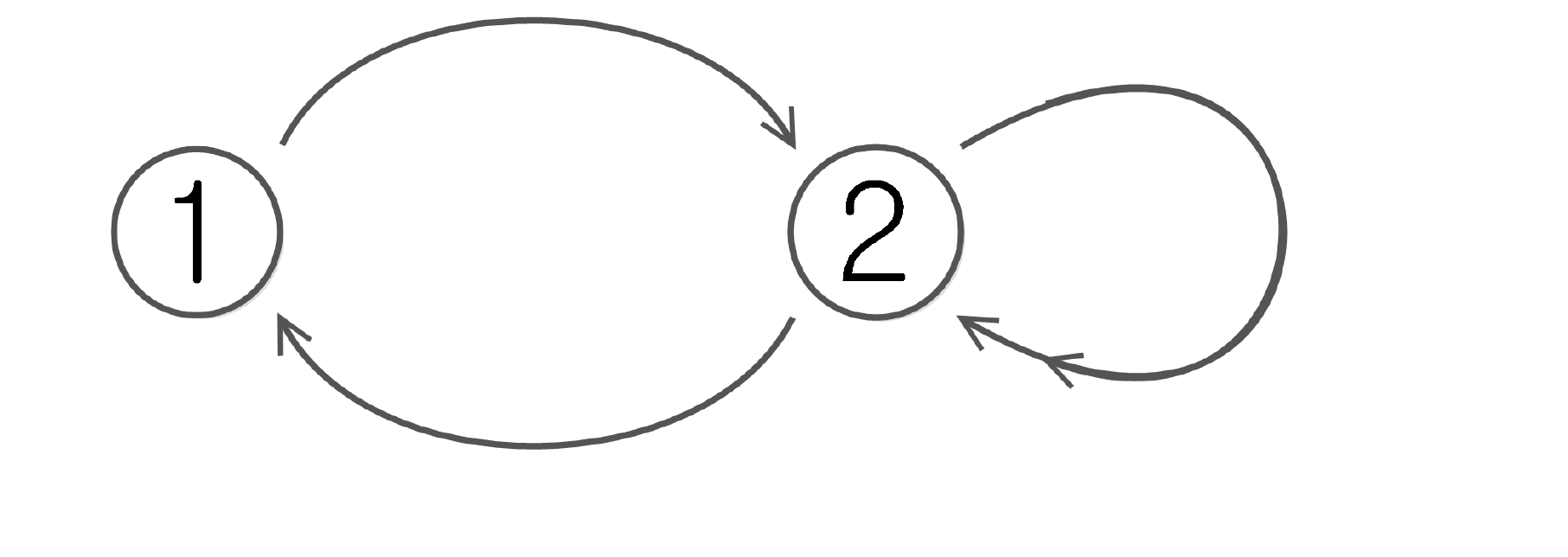}
\caption{The quiver diagram for the dual ABJM theory.}
\label{quiver-dualabjm}
\end{figure}
One can construct the gauge invariant operators (up to $F$-term conditions and taking $\Tr$ is understood)
\bea
\begin{array}{lll}
T^{(m)} X_{21}^{mk+n_2}X_{12}^{n_2} \phi_1^{n_3}\phi_2^{n_4}, &~~~~~\textrm{if} & m>0\, ,\\
T^{(m)} X_{21}^{n_1}X_{12}^{|m|k+n_1} \phi_1^{n_3}\phi_2^{n_4}, &~~~~~&m<0 \, .
\end{array}
\eea
Here $T^{(m)}$ is a monopole operator with monopole number $m$ and R-charge $m\Delta_m$.
 (When $m<0$, $T^{(m)}$ represents the anti-monopole operator with R-charge $-|m|\Delta_m$.)

Considering the flat directions of the matrix model, let us set the R-charges of the fields to be $R[X_{12}]=R[X_{21}]=\Delta_1$ and $R[\phi_i]=\Delta_{i+1}$. First study $m>0$ case. Let $n_1$ be the number of the chiral field $X_{21}$, and we have the following two equations.
\bea
n_1&=&mk+n_2, \nn \\
r&=&m\Delta_m +n_1 \Delta_1 +n_2 \Delta_1+n_3 \Delta_2+n_4 \Delta_3.
\eea
Then the number of operators with R-charge $r$ and monopole charge $m$ can be written as
\bea
\frac{\partial^2\psi}{\partial r\partial m}
&=&\int dn_1dn_2dn_3dn_4 \, \delta(n_1-mk-n_2) \nn \\
&\times& \delta(r-m\Delta_m -n_1 \Delta_1 -n_2 \Delta_1-n_3 \Delta_2-n_4 \Delta_3)
\nn \\
&=&\frac{1}{\Delta_2} \int dn_1 dn_4.
\eea
Note that the region of the surface integral should be bounded by 
\bea
&&n_1> mk,~~~~~ n_4>0, \nn \\
&&n_1< \frac{r-m(\Delta_m-k \Delta_1)}{2\Delta_1}-\frac{\Delta_3}{2\Delta_1}n_4. 
\eea
Similar calculations can be easily done for $m<0$. Using \eqref{opcrho}, one can obtain the eigenvalue density as 
\begin{displaymath}
\hat{\rho}(\hat{x})=\left\{
\begin{array}{lll}
\dfrac{1-\hat{x}(\Delta_m-k \Delta_1)}{2\Delta_1\Delta_2\Delta_3},&~~~& 
\dfrac{1}{\Delta_m-k\Delta_1}<\hat{x}<0,\\[5mm]
\dfrac{1-\hat{x}(\Delta_m+k \Delta_1)}{2\Delta_1\Delta_2\Delta_3},&~~~ &
0<\hat{x}<\dfrac{1}{\Delta_m+k\Delta_1}.
\end{array}
\right.
\end{displaymath}
The volume of the internal manifold dual to this theory can be calculated by integrating $\hat{\rho}(\hat{x})$ over $\hat{x}$ using (\ref{7-vol}) 
\be
\textrm{Vol}= -\frac{k \pi^4}{48(\Delta_m^2-k^2\Delta_1^2)\Delta_2\Delta_3}.
\ee
From the marginality of the superpotential, one should impose $2\Delta_1+\Delta_2+\Delta_3=2$. Then this volume is minimized at $\Delta_m=0, \Delta_1=1/2, \Delta_2=1/2$ to give $\pi^4/3k$, which is precisely the volume of $S^7/\mathbb{Z}_k$.

%%%%%%%%%%%%%%%%%%%%%%%%%%%%%%%%%%%%%%%%%%%%
\subsection{Chiral-like models with homogeneous dual manifold}
%%%%%%%%%%%%%%%%%%%%%%%%%%%%%%%%%%%%%%%%%%%%

\subsubsection{$Q^{1,1,1}/\mathbb{Z}_k$}

\begin{figure}
\centering
\includegraphics[scale=0.4,trim= 20 100 300 80,clip=true]{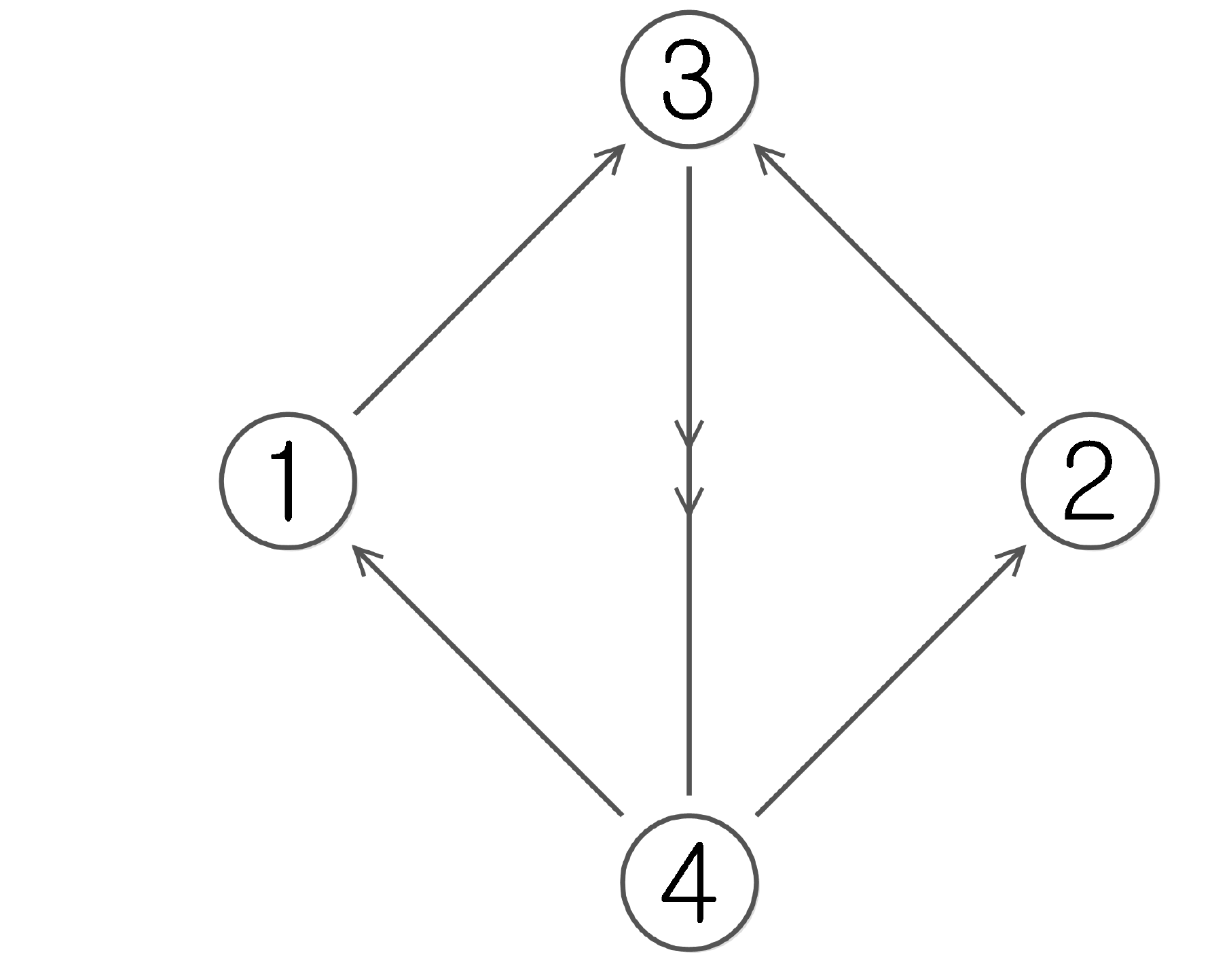}
\caption{The quiver diagram dual to $Q^{111}/\mathbb{Z}_k$}
\label{quiver-q111-diagonal}
\end{figure}

A 3-dimensional Chern-Simons matter theory dual to $AdS_4 \times Q^{111}/\mathbb{Z}_k$ was proposed in \cite{Franco:2008um,Franco:2009sp}. It has $U(N)^4$ gauge groups with  CS levels $(k,k,-k,-k)$ and 6 bifundamental fields. See the quiver diagram Fig.\ref{quiver-q111-diagonal}. The superpotential is given by
\be
W=\epsilon_{ij}\textrm{Tr} X_{41}X_{13}X_{34}^i X_{42} X_{23}X_{34}^j,~~~~~~~~i,j=1,2.
\ee
The gauge invariant operators are (The $SU(2)$ indices are suppressed and the superscripts denote exponents.)
\bea
\begin{array}{lll}
T^{(m)} X_{34}^{mk+n_1+n_2}X_{13}^{mk+n_1}X_{23}^{mk+n_2}X_{42}^{n_2}X_{41}^{n_1}, $~~~~~ \textrm{if} $m>0, \nn \\
T^{(m)} X_{34}^{|m|k+n_1+n_2}X_{13}^{n_1}X_{23}^{n_2}X_{42}^{|m|k+n_2}X_{41}^{|m|k+n_1}, $~~~~ $m<0.
\end{array}
\eea
Let $n_{i+2}$ to be the number of  fields $X_{34}^i$. One can set the R-charges of the fields to be $R[X_{41}]=R[X_{13}]=R[X_{34}^1]=\Delta_1, R[X_{42}]=R[X_{23}]=R[X_{34}^2]=\Delta_2$ by considering the flat directions. We assume $\Delta_1 \ge\Delta_2$ without loss of the generality. When $m>0$, we have the following relations
\bea
n_3+n_4&=&mk+n_1+n_2, \nn \\
r&=& m \Delta_m+ (n_3+mk+2 n_1)\Delta_1+(n_4+m k+2 n_2)\Delta_2.
\eea
The number of operators can be calculated as follows
\bea
\frac{\partial^2 \psi}{\partial r \partial m}
&=&\int dn_1 dn_2 dn_3 dn_4  \delta(n_3+n_4-mk-n_1-n_2) \nn \\
&\times& \delta\Big(r- m \Delta_m-(n_3+mk+2 n_1)\Delta_1-(n_4+m k+2 n_2)\Delta_2\Big),  \label{num-q111-diag} \\
&=&\frac{1}{\Delta_1-\Delta_2} \int dn_1 dn_2,
\eea
where the region is bounded by
\bea
&&n_1>0,~~~~~n_2>0,\nn \\[5mm]
&&n_1<\frac{r-m(\Delta_m+k(\Delta_1+2\Delta_2))}{2\Delta_1+\Delta_2}
-\frac{3\Delta_2}{2\Delta_1+\Delta_2}n_2, \nn \\[5mm]
&&n_1>\frac{r-m(\Delta_m+k(2\Delta_1+\Delta_2))}{3\Delta_1}
-\frac{\Delta_1+2\Delta_2}{3\Delta_1}n_2 \, . 
\eea
As a result, the eigenvalue density is
\begin{displaymath}
\hat{\rho}(\hat{x})=\left\{
\begin{array}{ll}
\tfrac{1-\hat{x}(\Delta_m-k(\Delta_1+2\Delta_2))}{3(\Delta_1-\Delta_2)\Delta_2(2\Delta_1+\Delta_2)},
&  \tfrac{1}{\Delta_m-k(\Delta_1+2\Delta_2)}<\hat{x}<\tfrac{1}{\Delta_m-k(2\Delta_1+\Delta_2)}, \\[5mm]
\tfrac{\Delta_1+\Delta_2-x(\Delta_m(\Delta_1+\Delta_2)-k(\Delta_1^2+\Delta_1\Delta_2+\Delta_2^2))}{3\Delta_1\Delta_2(2\Delta_1+\Delta_2)(\Delta_1+2\Delta_2)},
&  \tfrac{1}{\Delta_m-k(2\Delta_1+\Delta_2)}<\hat{x}<0,\\[5mm]
\tfrac{\Delta_1+\Delta_2-x(\Delta_m(\Delta_1+\Delta_2)+k(\Delta_1^2+\Delta_1\Delta_2+\Delta_2^2))}{3\Delta_1\Delta_2(2\Delta_1+\Delta_2)(\Delta_1+2\Delta_2)},
&  0<\hat{x}<\tfrac{1}{\Delta_m+k(2\Delta_1+\Delta_2)}, \\[5mm]
\tfrac{1-\hat{x}(\Delta_m+k(\Delta_1+2\Delta_2))}{3(\Delta_1-\Delta_2)\Delta_2(2\Delta_1+\Delta_2)},
& \tfrac{1}{\Delta_m+k(2\Delta_1+\Delta_2)}<\hat{x}< \tfrac{1}{\Delta_m+k(\Delta_1+2\Delta_2)}.
\end{array}
\right.
\end{displaymath}
By integrating $\hat{\rho}$ over $\hat{x}$, one can obtain the 7-dimensional internal space volume in terms of R-charges of the monopole operator and bifundamental fields
\be
\textrm{Vol}=\frac{k \pi^4(k^2(2\Delta_1+\Delta_2)^2(\Delta_1+2\Delta_2)^2
-\Delta_m^2(\Delta_1^2+\Delta_1\Delta_2+\Delta_2^2))}
{72\Delta_1\Delta_2(2\Delta_1+\Delta_2)(\Delta_1+2\Delta_2)(\Delta_m^2-k^2(2\Delta_1+\Delta_2)^2)(\Delta_m^2-k^2(\Delta_1+2\Delta_2)^2)}
\ee
Under the condition $\Delta_1+\Delta_2=\frac{2}{3}$ from the marginality of the superpotential, this volume is minimized at $\Delta_m=0,  \Delta_1=1/3$ to give correct volume of $Q^{111}/Z_k$, $\dfrac{\pi^4}{8k}$.
\\  \newline
 Now let us turn to the volume of 5-cycles. One can set $X_{ab}=0$ and count the operators without $X_{ab}$. 
First we set $X_{41}=0$ and count the number of operators with $n_1=0$ and $m>0$ case. The problem is reduced to integrating \eqref{num-q111-diag} with $n_1=0$ and calculate the length
\be
\frac{\partial^2\psi_{X_{41}}}{\partial r\partial m}=\frac{1}{\Delta_1-\Delta_2} \int dn_2
\ee
under the following conditions 
\be
n_2< \frac{r-m(\Delta_m+k(\Delta_1+2\Delta_2))}{3\Delta_2},~~~n_2>\frac{r-m(\Delta_m+k(2\Delta_1+\Delta_2))}{\Delta_1+2\Delta_2}.
\ee
As a result we obtain the matrix model quantity
\begin{displaymath}
\hat{\rho}(\hat{x}) S_{41}=\left\{
\begin{array}{lll}
\tfrac{1-\hat{x}(\Delta_m+k(\Delta_1-\Delta_2))}{3\Delta_2(\Delta_1+2\Delta_2)},
&&0<\hat{x}<\tfrac{1}{\Delta_m+k(2\Delta_1+\Delta_2)},\\[5mm]
\tfrac{1-\hat{x}(\Delta_m+k(\Delta_1+2\Delta_2))}{3\Delta_2(\Delta_1-\Delta_2)},
&&\tfrac{1}{\Delta_m+k(2\Delta_1+\Delta_2)}<\hat{x}<\tfrac{1}{\Delta_m+k(\Delta_1+2\Delta_2)},
\end{array}
\right.
\end{displaymath}
where we introduced a shorthand notation $S_{ab}=\hat{y}_b (\hat{x})- \hat{y}_a(\hat{x})+R(X_{ab})$.
We record this quantity for the remaining 5 fields.

$X_{13}=0$ : $n_1=0$ with $m<0$.
\begin{displaymath}
\hat{\rho}(\hat{x}) S_{13}=\left\{
\begin{array}{lll}
\tfrac{1-\hat{x}(\Delta_m-k(\Delta_1+2\Delta_2))}{3\Delta_2(\Delta_1-\Delta_2)},
&&\tfrac{1}{\Delta_m-k(\Delta_1+2\Delta_2)}<\hat{x}<\tfrac{1}{\Delta_m-k(2\Delta_1+\Delta_2)},\\[5mm]
\tfrac{1-\hat{x}(\Delta_m-k(\Delta_1-\Delta_2))}{3\Delta_2(\Delta_1+2\Delta_2)},
&&\tfrac{1}{\Delta_m-k(2\Delta_1+\Delta_2)}<\hat{x}<0.
\end{array}
\right.
\end{displaymath}

$X_{42}=0$ : $n_2=0$ with $m>0$
\begin{displaymath}
\hat{\rho}(\hat{x}) S_{42}=\left\{
\begin{array}{lll}
\tfrac{1-\hat{x}(\Delta_m-k(\Delta_1-\Delta_2))}{3\Delta_1(2\Delta_1+\Delta_2)},
&&0<\hat{x}<\tfrac{1}{\Delta_m+k(2\Delta_1+\Delta_2)},\\[5mm]
\tfrac{1-\hat{x}(\Delta_m+k(\Delta_1+2\Delta_2))}{(\Delta_1-\Delta_2)(2\Delta_1+\Delta_2)},
&&\tfrac{1}{\Delta_m+k(2\Delta_1+\Delta_2)}<\hat{x}<\tfrac{1}{\Delta_m+k(\Delta_1+2\Delta_2)}
\end{array}
\right.
\end{displaymath}

$X_{23}=0$ : $n_2=0$ with $m<0$
\begin{displaymath}
\hat{\rho}(\hat{x}) S_{23}=\left\{
\begin{array}{lll}
\tfrac{1-\hat{x}(\Delta_m-k(\Delta_1+2\Delta_2))}{(\Delta_1-\Delta_2)(2\Delta_1+\Delta_2)},
&&\tfrac{1}{\Delta_m-k(\Delta_1+2\Delta_2)}<\hat{x}<\tfrac{1}{\Delta_m-k(2\Delta_1+\Delta_2)},\\[5mm]
\tfrac{1-\hat{x}(\Delta_m-k(\Delta_1-\Delta_2))}{3\Delta_1(2\Delta_1+\Delta_2)},
&&\tfrac{1}{\Delta_m-k(2\Delta_1+\Delta_2)}<\hat{x}<0.
\end{array}
\right.
\end{displaymath}

$X_{34}^1=0$ : $n_3=0$
\begin{displaymath}
\hat{\rho}(\hat{x}) S_{34,1}=\left\{
\begin{array}{lll}
\tfrac{1-\hat{x}(\Delta_m-k(\Delta_1+2\Delta_2))}{3\Delta_2(2\Delta_1+\Delta_2)},
&& \tfrac{1}{\Delta_m-k(\Delta_1+2\Delta_2)}<\hat{x}<0,\\[5mm]
\tfrac{1-\hat{x}(\Delta_m+k(\Delta_1+2\Delta_2))}{3\Delta_2(2\Delta_1+\Delta_2)},
&&0<\hat{x}< \tfrac{1}{\Delta_m+k(\Delta_1+2\Delta_2)}.
\end{array}
\right.
\end{displaymath}

 $X_{34}^2=0$ : $n_4=0$
\begin{displaymath}
\hat{\rho}(\hat{x}) S_{34,2}=\left\{
\begin{array}{lll}
\tfrac{1-\hat{x}(\Delta_m-k(2\Delta_1+\Delta_2))}{3\Delta_1(\Delta_1+2\Delta_2)},
&& \tfrac{1}{\Delta_m-k(2\Delta_1+\Delta_2)}<\hat{x}<0,\\[5mm]
\tfrac{1-\hat{x}(\Delta_m+k(2\Delta_1+\Delta_2))}{3\Delta_1(\Delta_1+2\Delta_2)},
&&0<\hat{x}< \tfrac{1}{\Delta_m+k(2\Delta_1+\Delta_2)}.
\end{array}
\right.
\end{displaymath}
Using eq. (\ref{cycle-vol}), one can integrate this quantity and get the volume of the 5-cycles. In all cases, it gives $\dfrac{\pi^3}{4k}$
\footnote{In this model the chiral fields can be identified with the GLSM fields. See \cite{Franco:2008um} for example.} with $\Delta_m=0,  \Delta_1=1/3$
and agrees with geometric calculation \cite{Fabbri:1999hw}.
% The cone over $Q^{1,1,1}$ can be obtained by K\"ahler quotient $\mathbb{C}^6$ by $U(1)^2$ with the charges $()$.The chiral fields can be identified with GLSM fields \cite{Franco:2008um}.
From these computations we can predict the imaginary part of the eigenvalues in the matrix models.
\begin{eqnarray*}
%\begin{displaymath}
&&\hat{y}_1(\hat{x})-\hat{y}_4(\hat{x})\\
&&=\left\{
\begin{array}{ll}
-\Delta_1,
&  \frac{1}{\Delta_m-k(\Delta_1+2\Delta_2)}<\hat{x}
<
%\frac{1}{\Delta_m-k(2\Delta_1+\Delta_2)}
0
, \\[5mm]
%-\Delta_1,
%& \tfrac{1}{\Delta_m-k(2\Delta_1+\Delta_2)}<\hat{x}<0,\\[5mm]
-\Delta_1+\tfrac{\Delta_1(1-\hat{x}(\Delta_m+k(\Delta_1-\Delta_2)))(2\Delta_1+\Delta_2)}{\Delta_1+\Delta_2-\hat{x}(\Delta_m(\Delta_1+\Delta_2)+k(\Delta_1^2+\Delta_1\Delta_2+\Delta_2^2))},
&  0<\hat{x}<\frac{1}{\Delta_m+k(2\Delta_1+\Delta_2)}, \\[5mm]
\Delta_1+\Delta_2,
& \frac{1}{\Delta_m+k(2\Delta_1+\Delta_2)}<\hat{x}< \frac{1}{\Delta_m+k(\Delta_1+2\Delta_2)}.
\end{array}
\right.
%\end{displaymath}
\end{eqnarray*}

\begin{eqnarray*}
%\begin{displaymath}
&&\hat{y}_3(\hat{x})-\hat{y}_1(\hat{x})\\
&&=\left\{
\begin{array}{ll}
-\Delta_1+\tfrac{(1-\hat{x}(\Delta_m-k(\Delta_1+2\Delta_2)))(2\Delta_1+\Delta_2)}{1-\hat{x}(\Delta_m-k(\Delta_1+2\Delta_2))},
&  \frac{1}{\Delta_m-k(\Delta_1+2\Delta_2)}<\hat{x}<\frac{1}{\Delta_m-k(2\Delta_1+\Delta_2)}, \\[5mm]
-\Delta_1+\tfrac{\Delta_1(1-\hat{x}(\Delta_m-k(\Delta_1-\Delta_2)))(2\Delta_1+\Delta_2)}{\Delta_1+\Delta_2-\hat{x}(\Delta_m(\Delta_1+\Delta_2)-k(\Delta_1^2+\Delta_1\Delta_2+\Delta_2^2))},
&  \frac{1}{\Delta_m-k(2\Delta_1+\Delta_2)}<\hat{x}<0,\\[5mm]
-\Delta_1,
&  0<\hat{x}<
%\frac{1}{\Delta_m+k(2\Delta_1+\Delta_2)}
%, \\[5mm]
%-\Delta_1,
%& \frac{1}{\Delta_m+k(2\Delta_1+\Delta_2)}<\hat{x}< 
\frac{1}{\Delta_m+k(\Delta_1+2\Delta_2)}
.
\end{array}
\right.
%\end{displaymath}
\end{eqnarray*}

\begin{eqnarray*}
%\begin{displaymath}
&&\hat{y}_2(\hat{x})-\hat{y}_4(\hat{x})\\
&&=\left\{
\begin{array}{ll}
-\Delta_2,
&  \frac{1}{\Delta_m-k(\Delta_1+2\Delta_2)}<\hat{x}
%<\frac{1}{\Delta_m-k(2\Delta_1+\Delta_2)}
%, \\[5mm]
%-\Delta_2,
%&  \frac{1}{\Delta_m-k(2\Delta_1+\Delta_2)}<\hat{x}
<0,\\[5mm]
-\Delta_2+\tfrac{\Delta_2(1-\hat{x}(\Delta_m-k(\Delta_1-\Delta_2)))(\Delta_1+2\Delta_2)}{\Delta_1+\Delta_2-\hat{x}(\Delta_m(\Delta_1+\Delta_2)+k(\Delta_1^2+\Delta_1\Delta_2+\Delta_2^2))},
&  0<\hat{x}<\frac{1}{\Delta_m+k(2\Delta_1+\Delta_2)}, \\[5mm]
2\Delta_2,
& \frac{1}{\Delta_m+k(2\Delta_1+\Delta_2)}<\hat{x}< \frac{1}{\Delta_m+k(\Delta_1+2\Delta_2)}.
\end{array}
\right.
%\end{displaymath}
\end{eqnarray*}
Other variables such as $\hat{y}_4(\hat{x})-\hat{y}_3(\hat{x})$ and $\hat{y}_3(\hat{x})-\hat{y}_2(\hat{x})$ can be obtained easily using above results.
%%%%%%%%%%%%%%%%%%%%%%%%%%%%%%%%%%%%%%%%%%%%
\subsubsection{$Q^{1,1,1}/\mathbb{Z}_k$ and $Q^{2,2,2}/\mathbb{Z}_k$}\label{q111-box}
%%%%%%%%%%%%%%%%%%%%%%%%%%%%%%%%%%%%%%%%%%%%
In this subsection we study the Chern-Simons theory with $U(N)^4$ gauge groups, 8 bifundamental fields $X_{12}^i,X_{23}^i,X_{34}^i,X_{41}^i$ and $(k_1,k_2,-k_1,-k_2)$ CS levels.
\begin{figure}
\centering
\includegraphics[scale=0.40,trim= 20 150 350 80,clip=true]{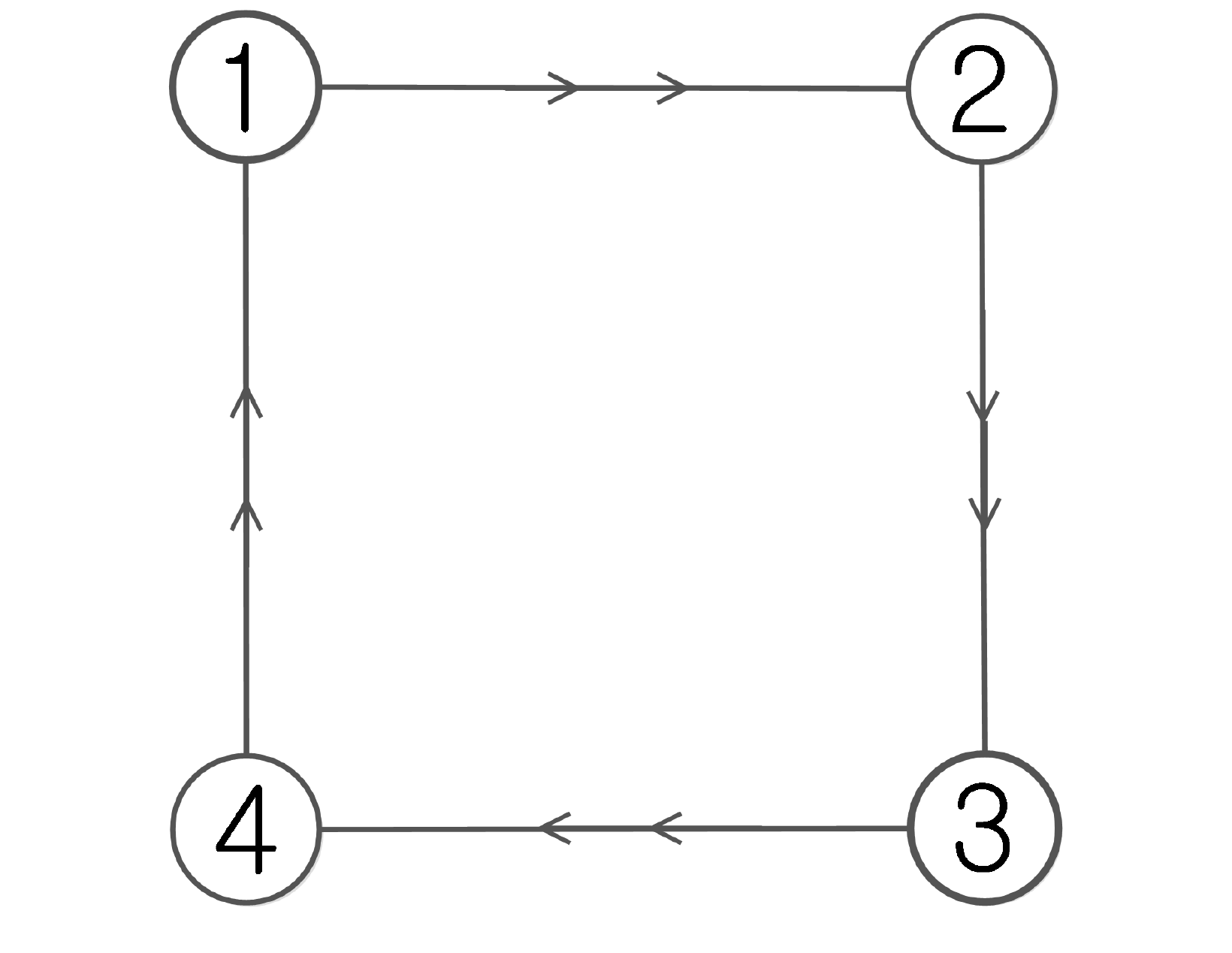}
\caption{The quiver diagram dual to $Q^{111}/\mathbb{Z}_k , ~Q^{222}/\mathbb{Z}_k $}
\label{quiver-q111}
\end{figure}
The superpotential is
\be
W=\epsilon_{ij}\epsilon_{kl}\textrm{Tr}X_{12}^i X_{23}^k X_{34}^j X_{41}^l
\ee
where $i$ is a $SU(2)$ index.
When $k_1=0,k_2=k$, this model is proposed to be dual to $AdS_4 \times Q^{111}/\mathbb{Z}_k$ \cite{Aganagic:2009zk}. With $k_1=k_2=k$, this theory is dual to $AdS_4 \times Q^{222}/\mathbb{Z}_k$ \cite{Franco:2009sp} and the operator counting is already done in \cite{Gulotta:2011aa}. For generic $k_1,k_2$ it is suggested that this quiver is dual to in-homogeneous examples $Y^{p,k}(B_4)$ with $B_4=\mathbb{CP}^1\times \mathbb{CP}^1$ \cite{Ueda:2008hx}. Having this generalization in mind, we count the operators for general CS levels.

The gauge invariant operators are
\bea
\begin{array}{lll}
T^{(m)} X_{12}^{m k_1+s} X_{23}^{m (k_1+k_2)+s}  X_{34}^{m k_2+s}  X_{41}^{s}, 
&~~~ \textrm{if} & m>0, \nn \\
T^{(m)} X_{12}^{|m| k_2+s} X_{23}^{s}  X_{34}^{|m|k_1+s}  X_{41}^{|m|(k_1+k_2)+s},&~~~\textrm{if} ~~&m<0,
\end{array}
\eea
where we have suppressed the $SU(2)$ index, and the superscripts represent the exponent. First consider $m>0$ case. Let $n_{12,1}$ be the number of the field $X_{12}^1$ in the operator. To account for $SU(2) \times SU(2)$ symmetry, we define
\be
n_i=n_{12,i}+n_{34,i} \, ,~~~~~~~n_{i+2}=n_{23,i}+n_{41,i} \, .
\ee 
From the form of the gauge invariant operators we obtain the following equations with $k_+=k_1+k_2$
\be
n_1+n_2=m k_+ +2s, ~~~~~n_3+n_4=m k_+ +2s.
\ee
Considering the flat directions one may set the R-charge of the fields to be $R[X_{12}^i]=R[X_{23}^i]=R[X_{34}^i]=R[X_{41}^i]=\Delta_{i}$. Then the R-charge of this gauge invariant operator becomes
\be
r= m \Delta_m +\Delta_1 n_1+\Delta_2 n_2+\Delta_1 n_3+\Delta_2 n_4.
\ee
Assuming that $\Delta_1 \ge \Delta_2$, the number of the gauge invariant operators with R-charge $r$ and monopole charge $m$ is
\bea
\frac{\partial^2 \psi}{\partial r \partial m} &=&\int dn_1 dn_2 dn_3 dn_4 ds 
 ~\delta(n_1+n_2+n_3+n_4-2m k_+-4s) \label{num-q111-box} \nn \\
 &\times&\delta(n_1+n_2-n_3-n_4) \delta(r- m \Delta_m -\Delta_1 n_1-\Delta_2 n_2-\Delta_1 n_3-\Delta_2 n_4) \nn \\
&=&\frac{1}{2(\Delta_1 -\Delta_2)}\int dn_1 dn_2 \, . 
\eea
The 2 dimensional integral can be obtained as the area on $(n_1,n_2)$ plane, bounded by
\bea
&&n_1>0,~~~~~~~~n_2>0, ~~~~~~~~n_2>m(k_1+k_2)-n_1, \nn\\
&&n_2<\frac{r-\Delta_m}{2\Delta_2}-\frac{\Delta_1+\Delta_2}{2\Delta_2} n_1,~~~
n_2>\frac{r-\Delta_m}{\Delta_1+\Delta_2}-\frac{2\Delta_1}{\Delta_1+\Delta_2} n_1 \, .
\eea
Calculation for $m<0$ case is also straightforward. One can easily obtain the eigenvalue densities using the formula (\ref{opcrho}).
\begin{displaymath}
\hat{\rho}(\hat{x})=\left\{
\begin{array}{lll}
\tfrac{1-\hat{x}(\Delta_m-2k_+\Delta_2)}{4(\Delta_1-\Delta_2)^2\Delta_2},
& & \tfrac{1}{\Delta_m-2k_+\Delta_2}<\hat{x}<\tfrac{1}{\Delta_m-k_+(\Delta_1+\Delta_2)},\\[5mm]
\tfrac{\Delta_1(1-\hat{x}(\Delta_m+2k_+\Delta_2))-\Delta_2(3-3\hat{x}\Delta_m+2k_+\hat{x}\Delta_2)}{4(\Delta_1-\Delta_2)^2\Delta_2(\Delta_1+\Delta_2)},
& & \tfrac{1}{\Delta_m-k_+(\Delta_1+\Delta_2)}<\hat{x}<\tfrac{1}{\Delta_m-2k_+\Delta_1},\\[5mm]
\tfrac{1-\hat{x}\Delta_m}{4\Delta_1\Delta_2(\Delta_1+\Delta_2)},
& &\tfrac{1}{\Delta_m-2k_+\Delta_1}<\hat{x}<\tfrac{1}{\Delta_m+2k_+\Delta_1},\\[5mm]
\tfrac{\Delta_1(1-\hat{x}(\Delta_m-2k_+\Delta_2))-\Delta_2(3-3\hat{x}\Delta_m-2k_+\hat{x}\Delta_2)}{4(\Delta_1-\Delta_2)^2\Delta_2(\Delta_1+\Delta_2)},
& & \tfrac{1}{\Delta_m+2k_+\Delta_1}<\hat{x}<\tfrac{1}{\Delta_m+k_+(\Delta_1+\Delta_2)},\\[5mm]
\tfrac{1-\hat{x}(\Delta_m+2k_+\Delta_2)}{4(\Delta_1-\Delta_2)^2\Delta_2},
& & \tfrac{1}{\Delta_m+k_+(\Delta_1+\Delta_2)}<\hat{x}<\tfrac{1}{\Delta_m+2k_+\Delta_2}.
\end{array}
\right.
\end{displaymath}
Then the volume can be expressed in terms of R-charge of the monopole operator and bifundamental fields $\Delta_m,\Delta_1,\Delta_2$  
\be
\label{vol-q111-box}
\textrm{Vol}=-\dfrac{\pi^4 k_+^3(\Delta_m^2+2k_+^2(\Delta_1^2+4\Delta_1\Delta_2+\Delta_2^2))}{24(\Delta_m^2-4k_+^2\Delta_1^2)(\Delta_m^2-4k_+^2\Delta_2^2)(\Delta_m^2-k_+^2(\Delta_1+\Delta_2)^2)}\ee 
This volume is extremized at $\Delta_m=0,\Delta_1=1/2$ and becomes
\be
\textrm{Vol}= \frac{\pi^4}{8k_+}
\ee
where we used the marginality condition of the superpotential $\Delta_1+\Delta_2=1$.
This gives $\textrm{Vol}(Q^{111}/\mathbb{Z}_k)=\pi^4/8k$ with $(k_1,k_2)=(0,k)$ and $\textrm{Vol}(Q^{222}/\mathbb{Z}_k)=\pi^4/16k$ with $(k_1,k_2)=(k,k)$, in consistence
with AdS/CFT.

Let us turn to the volume of the 5-cycles. By setting $X_{12}^1=0$, we have to count the number of operators with no $X_{12}^1$ field. The F-term condition of the superpotential gives $X_{34}^1=0$ also. So we should count the number of operators with $n_1=0$. After integrating eq. (\ref{num-q111-box}) with $n_1=0$, the number of operators without $X_{12}^1$ field is
\be
\frac{\partial^2 \psi_{X_{12}^1}}{\partial r \partial m}=\frac{1}{2(\Delta_1-\Delta_2)} \int dn_2,
\ee
where the region bounded by
\be
n_2> k_+ m,~~~~~n_2<\frac{r-m\Delta_m}{2\Delta_2},~~~~~n_2>\frac{r-m\Delta_m}{\Delta_1+\Delta_2}.
\ee
Using (\ref{opccy}) we have
\begin{displaymath}
\hat{\rho}(\hat{x}) S_{12,1}=\left\{
\begin{array}{lll}
\tfrac{1-\hat{x}(\Delta_m-2 k_+ \Delta_2)}{4 \Delta_2(\Delta_1-\Delta_2)},&&
 \tfrac{1}{\Delta_m-2k_+\Delta_2}<\hat{x}<\tfrac{1}{\Delta_m-k_+(\Delta_1+\Delta_2)},\\[5mm]
\tfrac{1-\hat{x}\Delta_m}{4 \Delta_2(\Delta_1+\Delta_2)}, &&
\tfrac{1}{\Delta_m-k_+(\Delta_1+\Delta_2)}<\hat{x}<\tfrac{1}{\Delta_m+k_+(\Delta_1+\Delta_2)},\\[5mm]
\tfrac{1-\hat{x}(\Delta_m+2 k_+ \Delta_2)}{4 \Delta_2(\Delta_1-\Delta_2)},&& 
\tfrac{1}{\Delta_m+k_+(\Delta_1+\Delta_2)}<\hat{x}<\tfrac{1}{\Delta_m+2k_+\Delta_2}.
\end{array}
\right.
\end{displaymath}
Integrating this quantity over $\hat{x}$, one can obtain the volume of the 5-cycles. 
%{\bf changed}
\be
\textrm{Vol}(\Sigma_{X_{12}^1})= \frac{\pi^3}{4} \int \hat{\rho}(\hat{x}) [\hat{y}_2 (\hat{x})- \hat{y}_1(\hat{x})+\Delta_1] \, d\hat{x}=\frac{\pi^3}{4 k_+}.
\ee
The imaginary part of the eigenvalues associated to the gauge group $U(N)_1$ and $U(N)_2$ are
\begin{eqnarray*}
%\begin{displaymath}
&&\hat{y}_2(\hat{x})-\hat{y}_1(\hat{x})\\
&&=\left\{
\begin{array}{lll}
-\Delta_2,
& & \tfrac{1}{\Delta_m-2k_+\Delta_2}<\hat{x}<\tfrac{1}{\Delta_m-k_+(\Delta_1+\Delta_2)},\\[5mm]
\tfrac{(1-\hat{x}(\Delta_m-2 k_+ \Delta_1))\Delta_2(\Delta_1+\Delta_2)}{\Delta_1(1-\hat{x}(\Delta_m+2k_+\Delta_2))-\Delta_2(3-3\hat{x}\Delta_m+2k_+\hat{x}\Delta_2)},
& & \tfrac{1}{\Delta_m-k_+(\Delta_1+\Delta_2)}<\hat{x}<\tfrac{1}{\Delta_m-2k_+\Delta_1},\\[5mm]
0,
& &\tfrac{1}{\Delta_m-2k_+\Delta_1}<\hat{x}<\tfrac{1}{\Delta_m+2k_+\Delta_1},\\[5mm]
\tfrac{(1-\hat{x}(\Delta_m+2 k_+ \Delta_1))\Delta_2(\Delta_1+\Delta_2)}{\Delta_1(1-\hat{x}(\Delta_m-2k_+\Delta_2))-\Delta_2(3-3\hat{x}\Delta_m-2k_+\hat{x}\Delta_2)},
& & \tfrac{1}{\Delta_m+2k_+\Delta_1}<\hat{x}<\tfrac{1}{\Delta_m+k_+(\Delta_1+\Delta_2)},\\[5mm]
-\Delta_2,
& & \tfrac{1}{\Delta_m+k_+(\Delta_1+\Delta_2)}<\hat{x}<\tfrac{1}{\Delta_m+2k_+\Delta_2}.
\end{array}
\right.
%\end{displaymath}
\end{eqnarray*}
%We set $X_{34,1}=0$. From the F-term condition we have $X_{12,1}=0$. Counting with $n_1=0$ gives the exactly same answer as before. 
%Setting  $X_{12,2}=0$, one can count the operator with $n_2=0$ and obtain the same 5-cycles volume. And the differences in y-values are added by $-\Delta_1+\Delta_2$ to the above results. $X_{34,2}=0$ means $n_2=0$.
\\ \newline
Next we set $X_{41}^1=0$. From the F-term condition $X_{41}^1 X_{23}^2=X_{41}^2X_{23}^1$, we have $X_{23}^1=0$ or $X_{41}^2=0$. In this case, there are two separate branches of  contributing operators. When $X_{23}^1=0$, we have $n_3=0$ and it gives the same result as before. When $X_{41}^2=0$, we should count operators with $s=0$ and $m>0$. This additional contribution amounts to 
\begin{displaymath}
\hat{\rho}(\hat{x}) S_{41,1}
|_{s=0}=\left\{
\begin{array}{lll}
0,&&
0<\hat{x}<\tfrac{1}{\Delta_m+2k_+\Delta_1},\\[5mm]
\tfrac{-1+\hat{x}(\Delta_m+2 k_+ \Delta_1)}{(\Delta_1-\Delta_2)^2}, &&
\tfrac{1}{\Delta_m+2 k_+\Delta_1}<\hat{x}<\tfrac{1}{\Delta_m+k_+(\Delta_1+\Delta_2)},\\[5mm]
\tfrac{1-\hat{x}(\Delta_m+2 k_+ \Delta_2)}{(\Delta_1-\Delta_2)},&& 
\tfrac{1}{\Delta_m+k_+(\Delta_1+\Delta_2)}<\hat{x}<\tfrac{1}{\Delta_m+2k_+\Delta_2}.
\end{array}
\right.
\end{displaymath}
Collecting these two contribution, the total eigenvalue distributions are
\begin{eqnarray*}
%\begin{displaymath}
&&\hat{y}_1(\hat{x})-\hat{y}_4(\hat{x})\\
&&=\left\{ 
\begin{array}{lll}
-\Delta_2,
& & \tfrac{1}{\Delta_m-2k_+\Delta_2}<\hat{x}<\tfrac{1}{\Delta_m-k_+(\Delta_1+\Delta_2)},\\[5mm]
\tfrac{(1-\hat{x}(\Delta_m-2 k_+ \Delta_1))\Delta_2(\Delta_1+\Delta_2)}{\Delta_1(1-\hat{x}(\Delta_m+2k_+\Delta_2))-\Delta_2(3-3\hat{x}\Delta_m+2k_+\hat{x}\Delta_2)},
& & \tfrac{1}{\Delta_m-k_+(\Delta_1+\Delta_2)}<\hat{x}<\tfrac{1}{\Delta-2k_+\Delta_1},\\[5mm]
0,
& &\tfrac{1}{\Delta_m-2k_+\Delta_1}<\hat{x}<\tfrac{1}{\Delta_m+2k_+\Delta_1},\\[5mm]
-3\tfrac{(1-\hat{x}(\Delta_m+2 k_+ \Delta_1))\Delta_2(\Delta_1+\Delta_2)}{\Delta_1(1-\hat{x}(\Delta_m-2k_+\Delta_2))-\Delta_2(3-3\hat{x}\Delta_m-2k_+\hat{x}\Delta_2)},
& & \tfrac{1}{\Delta_m+2k_+\Delta_1}<\hat{x}<\tfrac{1}{\Delta_m+k_+(\Delta_1+\Delta_2)},\\[5mm]
3\Delta_2,
& & \tfrac{1}{\Delta_m+k_+(\Delta_1+\Delta_2)}<\hat{x}<\tfrac{1}{\Delta_m+2k_+\Delta_2}.
\end{array}
\right.
%\end{displaymath}
\end{eqnarray*}
The volume of the 5-cycles corresponding to $X_{41}^1=0$ is then 
\be
\textrm{Vol}(\Sigma_{X_{41}^1})=\frac{\pi^3}{4 k_+}+\frac{\pi^3}{4 k_+}=\frac{\pi^3}{2 k_+}.
\ee
%Setting $X_{41,2}=0$ gives two cases. $X_{23,2}=0$ means $n_4=0$. $X_{41,1}=0$ means $S=0$ with $m>0$. 

Finally we set $X_{23}^1=0$ and obtain two branches: $X_{41}^1=0$ with $n_3=0$ and $X_{23}^2=0$ with $s=0, \,\, m<0$. The number of the operator without $X_{23}^1$ from the $s=0$ with $m<0$ are
\begin{displaymath}
\hat{\rho}(\hat{x}) S_{23,1}
|_{s=0}=\left\{
\begin{array}{lll}
\tfrac{1-\hat{x}(\Delta_m-2 k_+ \Delta_2)}{(\Delta_1-\Delta_2)},&& 
\tfrac{1}{\Delta_m-2k_+\Delta_2}<\hat{x}<\tfrac{1}{\Delta_m-k_+(\Delta_1+\Delta_2)},\\[5mm]
\tfrac{-1+\hat{x}(\Delta_m-2 k_+ \Delta_1)}{(\Delta_1-\Delta_2)^2}, &&
\tfrac{1}{\Delta_m-k_+(\Delta_1+\Delta_2)}<\hat{x}<\tfrac{1}{\Delta_m-2 k_+\Delta_1},\\[5mm]
0,&&
\tfrac{1}{\Delta_m-2k_+\Delta_1}<\hat{x}<0.
\end{array}
\right.
\end{displaymath}
Eigenvalues obtained from $n_3=0$ and $s=0, m<0$ are
\begin{eqnarray*}
%\begin{displaymath}
&&\hat{y}_3(\hat{x})-\hat{y}_2(\hat{x})\\
&&= \left\{
\begin{array}{lll}
3\Delta_2,
& & \tfrac{1}{\Delta_m+2k_+\Delta_2}<\hat{x}<\tfrac{1}{\Delta_m-k_+(\Delta_1+\Delta_2)},\\[5mm]
-3\tfrac{(1-\hat{x}(\Delta_m-2 k_+ \Delta_1))\Delta_2(\Delta_1+\Delta_2)}{\Delta_1(1-\hat{x}(\Delta_m+2k_+\Delta_2))-\Delta_2(3-3\hat{x}\Delta_m+2k_+\hat{x}\Delta_2)},
& & \tfrac{1}{\Delta_m-k_+(\Delta_1+\Delta_2)}<\hat{x}<\tfrac{1}{\Delta_m-2k_+\Delta_1},\\[5mm]
0,
& &\tfrac{1}{\Delta_m-2k_+\Delta_1}<\hat{x}<\tfrac{1}{\Delta_m+2k_+\Delta_1},\\[5mm]
\tfrac{(1-\hat{x}(\Delta_m+2 k_+ \Delta_1))\Delta_2(\Delta_1+\Delta_2)}{\Delta_1(1-\hat{x}(\Delta_m-2k_+\Delta_2))-\Delta_2(3-3\hat{x}\Delta_m-2k_+\hat{x}\Delta_2)},
& & \tfrac{1}{\Delta_m+2k_+\Delta_1}<\hat{x}<\tfrac{1}{\Delta_m+k_+(\Delta_1+\Delta_2)},\\[5mm]
-\Delta_2,
& & \tfrac{1}{\Delta_m+k_+(\Delta_1+\Delta_2)}<\hat{x}<\tfrac{1}{\Delta_m+2k_+\Delta_2}.
\end{array}
\right.
%\end{displaymath}
\end{eqnarray*}
For the other 5 fields, we can calculate similarly.  
\begin{displaymath}
\begin{array}{llll}
X_{34}^1=0 : & X_{12}^1=0 & \rightarrow & n_1=0, \\
X_{12}^2=0 : & X_{34}^2=0 & \rightarrow & n_2=0,\\
X_{34}^2=0 : & X_{12}^2=0 & \rightarrow & n_2=0,\\
X_{41}^2=0 : & X_{23}^2=0 & \rightarrow & n_4=0,\\
                         & X_{41}^1=0 & \rightarrow & s=0,m>0\\
X_{23}^2=0 : & X_{41}^2=0 & \rightarrow & n_4=0,\\
                         & X_{23}^1=0 & \rightarrow & s=0,m<0\\
\end{array}
\end{displaymath}
For example let us consider setting $X_{34}^1=0$. From the F-term condition we have $X_{12}^1=0$ also. This implies that we have to count the number of the operators with $n_1=0$.

As a result, the volume of the 5-cycles are
\bea
\textrm{Vol}(\Sigma_{X_{12}^i})&=&\textrm{Vol}(\Sigma_{X_{34}^i})=\frac{\pi^3}{4 k_+}, \nn \\
\textrm{Vol}(\Sigma_{X_{41}^i})&=&\textrm{Vol}(\Sigma_{X_{23}^i})=\frac{\pi^3}{2 k_+}.
\eea
To compare with the geometric computations we follow \cite{Gulotta:2011aa}. The cone over $Q^{1,1,1}$ can be obtained by the K\"ahler quotient of $\mathbb{C}^6$ by $U(1)^2$ with charges $(1,1,-1,-1,0,0)$ and $(1,1,0,0,-1,-1)$. Let us parametrize the coordinates on $\mathbb{C}^6$ with $(a_1,a_2,b_1,b_2,c_1,c_2)$. If we identify the chiral matter fields with the GLSM fields
\be
X_{12}^i=X_{34}^i=a_i,~~~X_{23}^i=b_i c_1,~~~ X_{41}^i=b_i c_2,
\ee
the volume of the 5-cycles can be written as
\bea
k\textrm{Vol}(\Sigma_{X_{12}^i})&=&k\textrm{Vol}(\Sigma_{X_{34}^i})=\textrm{Vol}(a_i),\nn \\
k\textrm{Vol}(\Sigma_{X_{23}^i})&=&\textrm{Vol}(b_i)+\textrm{Vol}(c_1),\nn \\ 
k\textrm{Vol}(\Sigma_{X_{41}^i})&=&\textrm{Vol}(b_i)+\textrm{Vol}(c_2).
\eea
It is consistent with the geometric computations \cite{Fabbri:1999hw}
\be
\textrm{Vol}(a_i)=\textrm{Vol}(b_i)=\textrm{Vol}(c_i)=\frac{\pi^3}{4}.
\ee
%%%%%%%%%%%%%%%%%%%%%%%%%%%%%%%
%%%%%%%%%%%%%%%%%%%%%%%%%%%%%%%%%%
\subsection{Chiral-like models from duals of $Y^{p,k}(B_4)$}
%%%%%%%%%%%%%%%%%%%%%%%%%%%%%%%%%%
%%%%%%%%%%%%%%%%%%%%%%%%%%%%%
Let us now consider inhomogeneous Sasaki-Einstein manifolds $Y^{p,k}(\mathbb{CP}^2)$ and their field theory duals. 
 The explicit form of the metric was constructed in \cite{Gauntlett:2004hh} as a higher-dimensional
generalization of their five-dimensional cousin $Y^{p,q}$. The seven-dimensional case
is analyzed in more detail by \cite{Martelli:2008rt} and here we provide a summary of its result which is relevant to us.

The metric of $Y^{p,k}(B_4)$ in the canonical form can be written as follows:
%\bea
%ds^2_7 = \frac{x}{4}d\tilde{s}^2_4 + 
%\frac{1}{4U(x)}d x^2 + q(x)(d\psi + A)^2 + \frac{w(x)}{16}(d \alpha + f(x)(d\psi + A))^2\label{y7metric} .
%\eea
%$B_4$ is the 4-dimensional base manifold and the various functions in the metric are
%\begin{align}
 % U(x)&=\frac{-3x^4+4x^3+\kappa}{3x^2} ,\\
%w(x)&=\frac{-2x^3+3x^2+\kappa}{3x^2}, \\
  %  q(x)&=\frac{-3x^4+4x^3+\kappa}{16(-2x^3+3x^2+\kappa)}, \\
  %f(x)&=\frac{x^3+\kappa}{-2x^3+3x^2+\kappa}.
%\end{align}
\bea
ds^2_7& = &(d\psi'+\sigma)^2+\frac{x}{4} d\tilde{s}^2_4 + \frac{1}{4U(x)}dx^2 + \frac{U(x)}{16}(d\phi - A)^2, \nn \\
\sigma &=& \frac{1}{4}A + \frac{1-x}{4}(d\phi - A), \nn \\
U(x)&=&\frac{-3x^4+4x^3+\kappa}{3x^2}, \label{metric-se7}
\eea
where $B_4$ is 4-dimensional K\"ahler-Einstein manifold and $\tfrac{1}{2}\textrm{d}A$ gives its K\"ahler two-form. Here
$\kappa$ is a constant satisfying $-1<\kappa<0$ and $x_1,x_2$ are two real zeroes of $U(x)$. Due to the positivity of the metric, $x$ should be in the range $x_1<x<x_2$. To avoid conical singularities $\kappa$, or equivalently $x_1, x_2$ should take certain discrete values. It is shown in \cite{Martelli:2008rt} that $x_1,x_2$ should satisfy 
\be
\frac{x_1(x_2-1)}{x_2(x_1-1)}=1-\frac{hp}{k}.\label{x1x2rel}
\ee
And more concretely, $x_1,x_2$ are real solutions to the following cubic equations
\bea
&&3p^3 x_1^3+2p^2(6b-5p)x_1^2+p(18b^2-28 p b+11p^2)x_1+4(3b^3+4p^2b-6pb^2-p^3)=0, \nn \\
&&3p^3x_2^3 +2p^2(p-6b)x_2^2+p(18b^2-8 p b+p^2)x_2+4b(3pb-3b^2-p^2)=0 .  \label{pqeq}
\eea
Here $b=k/h$, $h$ being the greatest common divisor of all Chern numbers for the base manifold $B_4$. $p,k$ are positive integers of our choice. 

Then the volume of the 7-dimensional manifold $Y^{p,k}(B_4)$ is
\be
\textrm{Vol}(Y^{p,k}(B_4))=\textrm{Vol}(B_4)\frac{\pi^2}{3 \cdot 2^7}\frac{x_2-x_1}{p(x_2-1)(1-x_1)}(x_2^3-x_1^3).\label{ypkcp2vol}
\ee
When $B_4$ is $\cp^2$ for instance, 
the volume of the 5-cycles are
\bea
\textrm{Vol}(\Xi_i) &=& \frac{9\pi^3}{64}\frac{x_2-x_1}{p(x_2-1)(1-x_1)}x_i^2 |x_i-1|~~~~~~i=1,2, \nn \\
\textrm{Vol}(\Theta_i) &=& \frac{3\pi^3}{64}\frac{x_2-x_1}{p(x_2-1)(1-x_1)}(x_2^2-x_1^2)~~~~~~i=1,2,3
\label{x1x2eq}
\eea
since $h=3,\textrm{Vol}(\cp^2)=9 \pi^2/2$.

%%%%%%%%%%%%%%%%%%%%%%%%%%%%%%%
\subsubsection{$B_4=\cp ^2$}
%%%%%%%%%%%%%%%%%%%%%%%%%%%%%%%%%%
%The dual geometry is an in-homogeneous Sasaki-Einstein manifold with $SU(3)\times U(1)$ isometry and usually dubbed $Y^{p,k}(\mathbb{CP}^2)$.
\begin{figure}
\centering
\includegraphics[scale=0.4,trim= 5 270 300 40,clip=true]{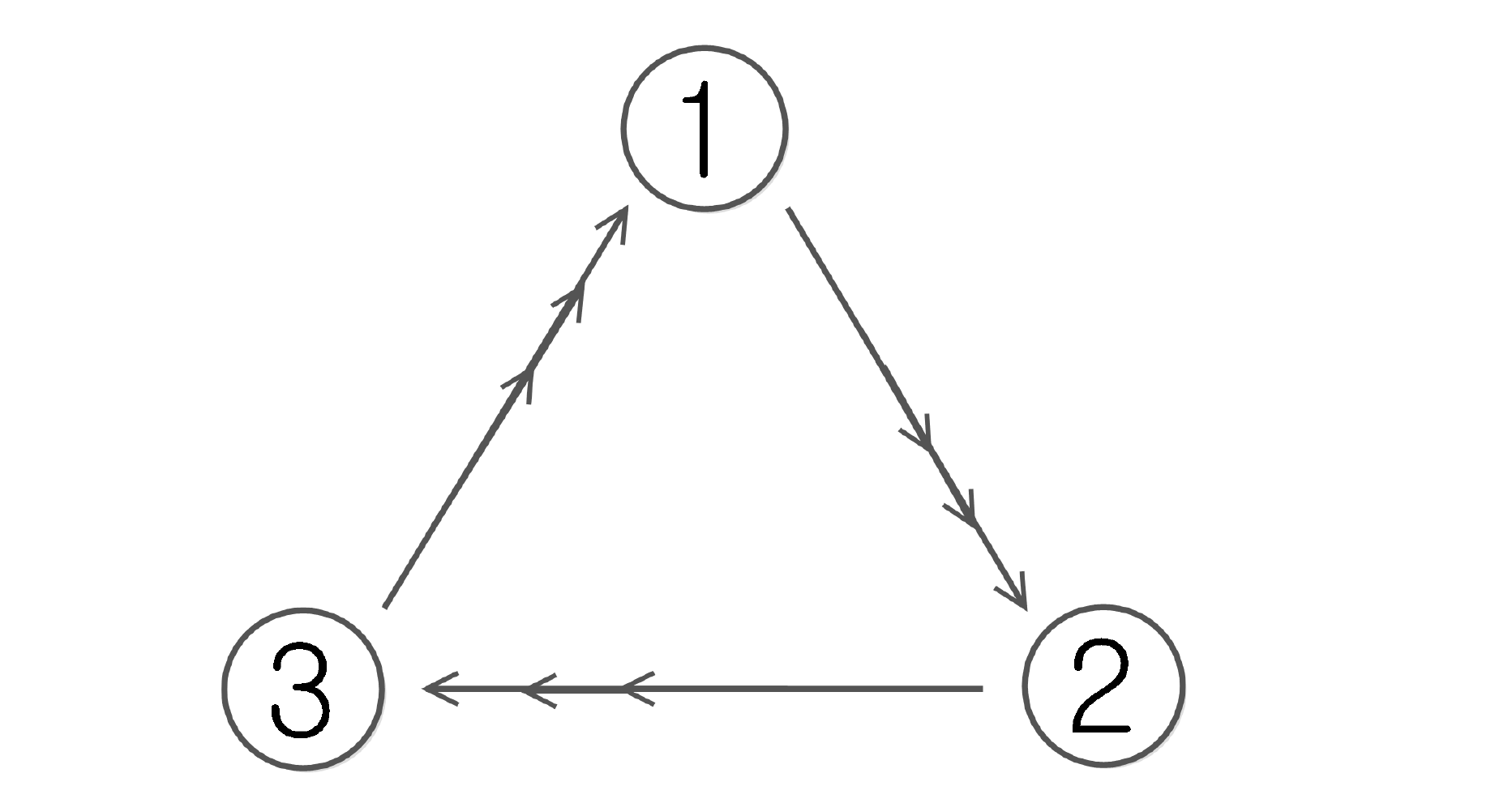}
\caption{The quiver diagram dual to $Y^{p,k}(\cp^2)$}
\label{quiver-ypkcp2}
\end{figure}

The Chern-Simons field theory of our interest here is chiral-like with quiver diagram Fig.\ref{quiver-ypkcp2}  \cite{Martelli:2008si}. 
We assign the CS levels as $(p,-2p+k,-k+p)$,\footnote{Note that the CS levels assignment in \cite{Martelli:2008si} is $(2p-k,k-p,-p)$. Our convention is equivalent to theirs up to permutation and overall sign flip. The theory is dual to $Y^{p,k}(\cp^2)$ in the range $3p/2 \le k \le 2p$.} and it can be shown that the vacuum moduli space is cone over $Y^{p,k}(\mathbb{CP}^2)$. The theory has $U(N)^3$ gauge group and 9 bifundamental fields $X_{12}^i, X_{23}^i$ and $X_{31}^i$ with superpotential $W=\epsilon_{ijk}\textrm{Tr}X_{12}^i X_{23}^j X_{31}^k$. Here $i=1,2,3$ account for a global 
$SU(3)$ symmetry.

Because the quiver diagram Fig.\ref{quiver-ypkcp2} does not have obvious symmetry other than flavor $SU(3)$, it is not clear what is the right R-charge assignment. But it turns out that from classical particle motion one can read off simple relations between the R-charge of various fields \cite{Kim:2010ja}. %\footnote{We used the Chern-Simons levels $(2p-k, k-p, -p)$.}, 
We consider a particle moving along the Reeb vector direction, i.e. we set  
 \be
 t= \nu \tau, ~~~\psi' = \omega \tau
 \ee
 and fix all the other coordinates constant, satisfies the particle equation of motion. Since this solution is BPS the classical quantities are exact at quantum level, and one can establish a mapping between the global charges in field theory and the angular momenta of the geodesic motion \cite{Kim:2010ja}. It turns out that $x=1$ orbits are dual to the operators without the monopole operator. $x=x_2$ orbits correspond to the operator with monopole operator $T^{(1)}$ and $x=x_1$ orbits to the operator with anti-monopole operator $T^{(-1)}$.
 Then one can determine the R-charges of the gauge invariant operators using geometric quantities $p, k, x_1$ and $x_2$, as follows (The $SU(3)$ indices are suppressed and the superscripts here and below represent the exponent.)
 \be
 R[T^{(1)}X_{12}^{k-p}X_{31}^p]=\frac{2k}{3x_2}
 , ~~~R[T^{(-1)}X_{12}^{2p-k}X_{23}^p]=\frac{2(3p-k)}{3x_1}.\label{rcharge}
 \ee
  Note that we can easily see that these assignments are compatible with the exact marginality of superpotential, i.e. $R[X_{12}]+R[X_{23}]+R[X_{31}]=2$, using \eqref{x1x2rel}.

Now we may consider counting of 
 gauge invariant operators. In general they are expressed as 
\bea
\begin{array}{lll}
T^{(m)} X_{12}^{m k_1+s} X_{23}^{m (k_1+k_2)+s}  X_{31}^{s}, 
&~~~ \textrm{if} & m>0, \nn \\
T^{(m)} X_{12}^{s} X_{23}^{|m|k_2+s}  X_{31}^{|m|(k_2+k_3)+s} ,&~~~~~&m<0 . 
\end{array}
\eea
The subsequent calculations are in fact already performed in \cite{Gulotta:2011aa}, and here we will illustrate that the density function indeed gives rise to the volume formula
\eqref{ypkcp2vol}. 

%From eq.(5.5) of \cite{Gulotta:2011aa}, one can write down the volume function in terms of the monopole operator R-charge $\Delta_m$ as follows. 
%\bea
%&&
%\textrm{Vol}(Y^{p,k}(\cp^2)) \nn \\
%&&
%=\tfrac{27\pi^4}{16} \tfrac{(\kp+\km)(4 \km^2 \kp^2+6\km^2 \kp \Delta_m
%-6 \km \kp^2 \Delta_m +3\km^2 \Delta_m^2-3\km\kp\Delta_m^2+3\kp^2\Delta_m^2)}
%{(2\km-3\Delta_m)^3(2\kp+3\Delta_m)^3}\label{vol} \, , 
%\eea
 %where  $\km \equiv k_1-k_2 =3p-k, ~~\kp \equiv k_1-k_3=k$ and we have fixed $R[X_{12}^i]=R[X_{23}^i]=R[X_{31}^i]=\frac{2}{3}$.
From eq.(5.5) of \cite{Gulotta:2011aa}, one can write down the volume function in terms of the R-charges of the bifundamental fields and the monopole operator. Extremizing this with respect to the R-charges of the bifundamental fields gives $R[X_{12}^i]=R[X_{23}^i]=R[X_{31}^i]=\frac{2}{3}$. As the result, the volume is now a function of the monopole operator R-charge $\Delta_m$. 
\bea
&&
\textrm{Vol}(Y^{p,k}(\cp^2)) \nn \\
&&
=\tfrac{27\pi^4}{16} \tfrac{(\kp+\km)(4 \km^2 \kp^2+6\km^2 \kp \Delta_m
-6 \km \kp^2 \Delta_m +3\km^2 \Delta_m^2-3\km\kp\Delta_m^2+3\kp^2\Delta_m^2)}
{(2\km-3\Delta_m)^3(2\kp+3\Delta_m)^3}\label{vol} \, , 
\eea
 where  $\km \equiv k_1-k_2 =3p-k, ~~\kp \equiv k_1-k_3=k$.
  We now demand this quantity be minimized with respect to $\Delta_m$, and obtain 
 \footnote{We note that this equation can be also obtained when we extremize Eq.(7.10) in \cite{Berenstein:2011dr} with respect to $R_+$.}
 \bea
 &&18(\km^3+\kp^3)\Delta_m^3-3(\km^2-\kp^2)(\km^2-16\km\kp+\kp^2)\Delta_m^2 \nn \\
 &&-8\km\kp(\km+\kp)(\km^2-7\km\kp+\kp^2)\Delta_m-8 \km^2\kp^2(\km^2-\kp^2)=0. \label{deq}
 \eea
 For given $k_1,k_2,k_3$, this is a cubic equation for $\Delta_m$. To obtain the volume, one should substitute the solution for $\Delta_m$ into \eqref{vol}. 
 Now we would like to show that this result is always the same as \eqref{ypkcp2vol}, with $x_1,x_2$ satisfying \eqref{pqeq}.
 At first sight they look different, but we can show they are equivalent. The R-charge of BPS particle solutions \eqref{rcharge} lead us to define 
$x_1$ and $x_2$
  %\footnote{We found this AdS/CFT mapping in studying the geodesic motions in $AdS_4 \times Y^{p,k}(\cp^2)$ background reviewed in the \ref{geodesic}. With the above R-charge assignments, we can obtain these from the eq.(\ref{rcharge}).  See \cite{Kim:2010ja} for the details.} 
  as follows
 \bea
 x_1&=& \frac{2\km}{2\km-3\Delta_m}, \nn \\
 x_2&=& \frac{2\kp}{2\kp+3\Delta_m}.
 \eea
Then one can show that provided $\Delta_m$ satisfies \eqref{deq}, $\eqref{vol}$ can be re-expressed in terms of $x_1, x_2$
\be
\textrm{Vol}(Y^{p,k}(\cp^2))=\frac{3\pi^4}{256}\frac{x_2-x_1}{p(1-x_1)(x_2-1)}(x_2^3-x_1^3). 
\ee
Of course this is the same expression as ($\ref{ypkcp2vol}$) with $B_4=\cp^2$.
One can also check that the equation (\ref{deq}) leads to the cubic equations (\ref{pqeq}). So the operator counting method reproduces the volume of inhomogeneous 7 dimensional geometry $Y^{p,k}(\cp^2)$.
%\bea
%&&3p^3 x_1^3+2p^2(6b-5p)x_1^2+p(18b^2-28 p b+11p^2)x_1+4(3b^3+4p^2b-6pb^2-p^3)=0, \nn \\
%&&3p^3x_2^3 +2p^2(p-6b)x_2^2+p(18b^2-8 p b+p^2)x_2+4b(3pb-3b^2-p^2)=0.
%\eea

With $\hat{y}_{i+1}(\hat{x})-\hat{y}_i(\hat{x})$ (eq.(5.6), (5.7) in \cite{Gulotta:2011aa}), the volumes of the five-cycles can be computed as
\bea
\textrm{Vol}(\Sigma_{X_{23}^i}) &=& \frac{\pi^3}{4} \int d\hat{x}\hat{\rho}(\hat{x})
%(\hat{y_3}(\hat{x})-\hat{y_2}(\hat{x})+\frac{2}{3})
S_{23,i} =\frac{3\pi^3}{64}
\frac{(x_1-x_2)^2(x_1+x_2)}{p(1-x_1)(x_2-1)}, \nn \\
\textrm{Vol}(\Sigma_{X_{31}^i}) &=& \frac{\pi^3}{4} \int d\hat{x}\hat{\rho}(\hat{x})
%(\hat{y_1}(\hat{x})-\hat{y_3}(\hat{x})+\frac{2}{3})
S_{31,i}=\frac{3\pi^3}{64}
\frac{(x_1-x_2)(x_1^2+(2-3x_2)x_2^2)}{p(1-x_1)(x_2-1)}, \nn \\
\textrm{Vol}(\Sigma_{X_{12}^i}) &=& \frac{\pi^3}{4} \int d\hat{x}\hat{\rho}(\hat{x})
%(\hat{y_2}(\hat{x})-\hat{y_1}(\hat{x})+\frac{2}{3})
S_{12,i}=\frac{3\pi^3}{64}
\frac{(x_1-x_2)(-2x_1^2+3x_1^3-x_2^2)}{p(1-x_1)(x_2-1)}. 
\eea
%The volumes of the five-cycles in $Y^{p,k}(\cp^2)$ are 
%\bea
%\textrm{Vol}(\Xi_i) &=& \frac{9\pi^3}{64}\frac{x_2-x_1}{p(1-x_1)(x_2-1)}x_i^2 |x_i-1|~~~~~~i=1,2, \nn \\
%\textrm{Vol}(\Theta_i) &=& \frac{3\pi^3}{64}\frac{x_2-x_1}{p(1-x_1)(x_2-1)}(x_2^2-x_1^2)~~~~~~i=1,2,3.
%�\eea
Note that this volume is related to that obtained in the $Y^{p,k}(\cp^2)$ geometry.
\bea
\textrm{Vol}(\Sigma_{X_{23}^i}) &=& \textrm{Vol}(\Theta_i), \nn \\
 \textrm{Vol}(\Sigma_{X_{12}^i}) &=& \textrm{Vol}(\Theta_i)+\textrm{Vol}(\Xi_1), \nn \\
 \textrm{Vol}(\Sigma_{X_{31}^i}) &=& \textrm{Vol}(\Theta_i)+\textrm{Vol}(\Xi_2).
\eea

\subsubsection{$B_4=\cp^1 \times \cp^1$}
There has been an attempt to construct the gauge theory dual of $AdS_4 \times Y^{p,k}(\cp^1 \times \cp^1)$ background in \cite{Ueda:2008hx}\cite{Closset:2012ep}. This theory has the same quiver diagram and the superpotential as the models considered in Sec.\ref{q111-box}, but has more general CS levels $k_1=k-p, k_2=-p$. The authors of \cite{Ueda:2008hx} used a dimer model technique to obtain this model, but reported that one of the toric vectors for vacuum moduli space is outside a convex polytope. We can see that the operator counting method also becomes problematic. When one tries to extremize \eqref{vol-q111-box}, one obtains
\be
\Delta_m(5(k_1+k_2)^2+\Delta_m^2)=0.
\ee
Obviously $\Delta_m=0$ and the minimized volume will never be associated with a cubic equation like \eqref{pqeq}.
%\cite{Tomasiello:2010zz}
%%%%%%%%%%%%%%%%%%%%%%%%%%%%%%%%%%%%%%%%%
\section{Discussions}\label{discussion}
%%%%%%%%%%%%%%%%%%%%%%%%%%%%%%%%%%%%%%%%%
The AdS/CFT correspondence is a fascinating arena where quantum field theory and algebraic geometry are deeply inter-connected with each other. For the case of $D=4$ superconformal field theories with $AdS_5\times Y_5$ duals, the $a$-maximization theorem can be re-interpreted as volume minimization of Sasaki-Einstein manifold $Y_5$ \cite{Intriligator:2003jj}\cite{Martelli:2005tp}. %\cite{Martelli:2006yb}. 
The extension to seven-dimensional Sasaki-Einstein manifolds is straightforward, while the quantitative description of dual $D=3$ superconformal field theory was missing until recently. It turns out that the free energy $F$ on $S^3$ is an analogous quantity to the central charge $a$, and the F-theorem \cite{Gulotta:2011si} tells us what characterizes the IR fixed point of ${\cal N}=2$ gauge field theories.  

But M2-brane theories on generic singular Calabi-Yau manifolds are not fully understood yet. There is no first-principle derivation of the Chern-Simons gauge theory in general, and it is often the case that the only justification of a field theory dual proposal is the agreement of vacuum moduli space with (the cone of) seven-dimensional internal space which is Sasaki-Einstein. Thus the full-fledged quantum computation of partition function using localization technique, if applicable, should be essential in establishing the duality relation. 

It turns out that, at least for vector-like models the localization formula is amenable to semiclassical approximation in large-$N$ limit and the result at leading order correctly reproduces the volume of $d=7$ internal manifold. However, there are also several examples with chiral-like matter representations. To the best our knowledge the chirality of dual Chern-Simons model is not associated with the geometric data: for instance, there are chiral-like as well as vector-like dual models for $AdS_4\times Q^{1,1,1}$ background.

For chiral-like models in the large-$N$ limit, application of the technique in \cite{Herzog:2010hf} does not lead to the expected $N^{3/2}$ behavior of the free energy. In particular, the roots of saddle point equations do not seem to converge on a smooth cut \cite{Jafferis:2011zi}. Since both analytic and numerical approaches fail, the duality proposal of chiral-like models, e.g. $M^{1,1,1}$ or $Y^{p,k}(\mathbb{CP}^2)$, remain un-confirmed. 

There appeared two suggestions which might help overcome this impasse. One is the relationship between the root distribution of matrix model for partition function and the counting of chiral operators in gauge theory. The other is the symmetrization prescription proposal, which effectively turns the saddle point equation of the matrix model into that of a vector-like one. It is illustrated that for chiral-like duals of  $Q^{1,1,1}, M^{1,1,1}$ at least this prescription leads to a correct result for partition function \cite{Amariti:2011uw}\cite{Gang:2011jj}. 

The aim of this article was to check if any of these prescriptions can be applied to more nontrivial models, especially in-homogeneous models $Y^{p,k}(B_4)$. Unlike homogeneous examples, their volume is given as a fairly complicated irrational number and an agreement would establish a very strong evidence that the conjecture is correct. As it turns out, the operator counting method gives the correct volume formula after extremization. But the free energy from the symmetrized integrand is not extremized by a symmetrized eigenvalue distribution.

It is thus clear that for a better understanding of the quiver matrix models the operator counting provides very useful information. Of course, the operator counting is not really a quantum computation: It is more like an alternative way to extract geometrical data from the quiver diagram and superpotential. However, one might still use the eigenvalue distribution functions reported in \cite{Gulotta:2011aa} or in this article as a hint, and try to find alternative saddle point equations or quiver diagrams for chiral-like models. It will be very interesting if such matrix models can be reverse-engineered.

%\cite{Benini:2011cma,Closset:2012ep}
%%%\section*{Acknowledgments}
%%%%%%%%%%%%%%%%%%%%%%%%%%%%%%%%%%%%%%%%%%%%
\acknowledgments
We thank A. Amariti, D. Berenstein, C. Klare, and M. Siani for comments on the first version of the paper. 
This work was supported by a post-doctoral fellowship grant from Kyung Hee University (KHU-20110694).
The research of NK is supported by the National Research Foundation of Korea (NRF) funded by the Korean Government (MEST) with grant No. 2009-0085995, 2010-0023121, and also through the Center for Quantum Spacetime (CQUeST) of Sogang University with grant No. 2005-0049409. NK also gratefully acknowledges the hospitality of the Institute for Advanced Study, where part of this work was completed. 
%%%%%%%%%%%%%%%%%%%%%%%%%%%%%%%%%%%%%%%%
\bibliography{opc-JHEP}
\end{document}